\documentclass[12pt,preprint]{aastex}

\slugcomment{.}

\shorttitle{}
\shortauthors{Lieu}
\begin{document}

\title{The effect of long range gravitational perturbations on the
first acoustic peak of the cosmic microwave background}

\author{Richard Lieu\altaffilmark{1}}

\affil{Department of Physics, University of Alabama at Huntsville,
    Huntsville, AL 35899}

\begin{abstract}
In the standard cosmological model, the temperature anisotropy of
the cosmic microwave background is interpreted as variation in the
gravitational potential at the point of emission, due to the emitter
being embedded in a region ${\cal C}$ of over- or under-density spanning the
length (or size) scale $\lambda$
on which the anisotropy is measured.  If the Universe is
inhomogeneous, however, similar density contrasts of size $\lambda$
are also located everywhere surrounding ${\cal C}$.  Since
they are superposition states of many 
independent Fourier modes with no preferred direction,
such primordial clumps and voids should not be configured according to some 
prescribed spatial pattern.
Rather, they
can randomly trade spaces with each other while preserving the
Harrison-Zeldovich
character of the matter spectrum.
The outcome is an {\it extra}
perturbation of the potential when averaged 
over length $\lambda$ at the emitter, 
and consequently an additional anisotropy on
the same scale, which has apparently been overlooked.  Unlike the 
conventional application of the Sachs-Wolfe effect to the WMAP observations,
this extra effect is {\it not} scale independent over the $P(k) \sim k$ part
of the matter spectrum, but increases towards smaller lengths, as $\sqrt{k}$.
The consequence is a substantial revision of the currently advertised
values of the key cosmological parameters, {\it unless} one postulates
a more rapid decrease in the gravitational force with distance than
that given by the inverse-square law.
\end{abstract}

\vspace{3mm}

\noindent
{\bf 1. Introduction and preliminaries}

In their seminal paper, Sachs and Wolfe (1967) discussed
the effect of an inhomogeneous Universe at
the decoupling epoch on the temperature anisotropy of the cosmic
microwave background (CMB).  They considered random density perturbations
$\delta\rho/\rho$ over some characteristic length scale $\lambda$ -
a phenomenon
which was subsequently conjectured to have
originated from scale invariant fluctuations in the
primordial plasma (Harrison 1970,
Zeldovich 1972, Peebles \& Yu 1970).
This behavior, and its manifestation as
CMB temperature anisotropy, has been investigated in detail both
theoretically
(Peebles 1982, Bond \& Efstathiou 1984 and 1987) and observationally
(Bennett et al 2003, Page
et al 2003, and Spergel et al 2003 provided results from the latest
all-sky measurements by WMAP).  
The purpose of the present work is to point out that the conversion
from matter inhomogeneity to CMB anisotropy, adopted
by conventional interpretation of the WMAP TT
cross correlation data, is a restricted version of the Sachs-Wolfe
effect which does not take into account all contributions (to
the anisotropy) of comparable magnitude.

Let us first set up the necessary preliminary framework.  At
a spatial corrdinate ${\bf r}$ and referring all physical quantities to
their values at the present epoch where the expansion parameter is
$a_0 = 1$ let the matter density be $\rho({\bf r})=\rho_0+\delta\rho({\bf r})$
with the perturbation term having zero spatial average over some large volume,
i.e. $\langle\delta\rho({\bf r})\rangle=0$.  The Fourier transform
of $\rho$, and its inverse transform, are
\begin{equation}
 \tilde\rho({\bf k})=\int d^3{\bf r}\,e^{-i{\bf k\cdot r}}\rho({\bf
r}),\qquad
 \rho({\bf r})=\int \frac{d^3{\bf k}}{(2\pi)^3}\,e^{i{\bf k\cdot
r}}\tilde\rho({\bf k}),
\end{equation}
respectively.  The deviation in the gravitational potential at
${\bf r}$, due to the inhomogeneous Universe, is given by
\begin{equation}
\delta\Phi({\bf r})=-G\int d^3{\bf r}'\frac{\delta\rho({\bf r}')}{|{\bf
r}-{\bf r}'|}.
\end{equation}
We shall return to Eq. (2) soon.  For the moment, let us examine
$\delta\tilde\rho({\bf k})$, because the power spectrum $P(k)$ is characterized
by an equation which involves $\delta\tilde\rho$:
\begin{equation}
\langle\delta\tilde\rho({\bf k})\delta\tilde\rho({\bf k}')\rangle=
 \rho_0^2P(k)(2\pi)^3\delta({\bf k}+{\bf k}').
\end{equation}
For the Harrison-Zeldovich (HZ) spectrum, $P(k) \sim k$.

We now define a smoothing function 
(or filter) $W_\lambda({\bf r})$ corresponding to
any chosen scale $\lambda$, satisfying $W_\lambda({\bf r})\approx1$ for
$r\ll\lambda$ and $W_\lambda({\bf r})\approx0$ for $r\gg\lambda$.  The
effective smoothing volume $V_{\lambda}$ is:
\begin{equation}
 V_\lambda=\int d^3{\bf r}\,W_\lambda({\bf r}).
\end{equation}
For example, a convenient choice is $W_\lambda(r)=e^{-r^2/2\lambda^2}$ for
which $V_\lambda=(2\pi)^{3/2}\lambda^3$ and $\tilde W_\lambda({\bf
k})=V_\lambda e^{-\lambda^2 k^2/2}$.
Then the smoothed mass over length scale $\lambda$ will be
\begin{equation}
\delta M_\lambda({\bf r})=\int d^3{\bf r}'\,W_\lambda({\bf r}-{\bf
r}')\delta\rho({\bf r}'),
\end{equation}
or equivalently
\begin{equation}
\delta\tilde M_\lambda({\bf k})=\tilde W_\lambda({\bf
k})\delta\tilde\rho({\bf k}).
\end{equation}
The averages of $M_{\lambda}$ are $\langle\delta M_\lambda({\bf r})\rangle=0$
and
\begin{equation}
\langle\delta M_\lambda({\bf r})\delta M_\lambda({\bf r}')\rangle=\int
\frac{d^3{\bf k}}{(2\pi)^3}\,e^{i{\bf k\cdot (r-r}')}|\tilde
W_\lambda({\bf k})|^2\rho_0^2P(k).
\end{equation}
Likewise for the smoothed density $\delta\rho_{\lambda}$ we have
$\langle\delta\rho_{\lambda} ({\bf r})\rangle =0$ and
$$
\langle\delta\tilde\rho_\lambda({\bf k})\delta\tilde\rho_\lambda({\bf
k}')\rangle=
 \rho_0^2e^{-\lambda^2k^2}P(k)(2\pi)^3\delta({\bf k}+{\bf k}').
$$
In particular when $P(k)=Ak^n$, 
the mean square value of the
smoothed mass can be evaluated by the saddle-point approximation.  The $k$
integral is dominated by the region near
$k=k_\lambda=\sqrt{1+\frac{n}{2}}/\lambda$, to give
\begin{equation}
 \langle (\delta
M_\lambda)^2\rangle=c_nM_{\lambda}^2k_\lambda^3P(k_\lambda),
\end{equation}
where $M_{\lambda}=\rho_0 V_\lambda$ and $c_n$ is a numerical constant.
Here-and-after we shall define the standard deviation in 
$\delta M_{\lambda} ({\bf r})$ as
\begin{equation}
\delta M_{\lambda}^{{\rm rms}} = 
\langle(\delta M_{\lambda})^2\rangle^{\frac{1}{2}}.
\end{equation}
A note of caution is already in order here.  Although the r.m.s. mass
and density over some length scale $\lambda$ at position
${\bf r}$ concern the distribution of matter {\it local} to ${\bf r}$,
the same cannot be said about r.m.s. values of 
the potential fluctuation $\delta\Phi ({\bf r})$, Eq. (2).  Owing
to the long range nature of the gravitational force, density contrasts
of size $\lambda$ but spreading over distances far greater than
$\lambda$ can also contribute towards $\delta\Phi_{\lambda}^{{\rm rms}}$.
 
\vspace{2mm}

\noindent
{\bf 2. CMB temperature anisotropy from primordial density
fluctuations: any missing component?}

If at some point ${\bf r}$ on the last scattering surface there is a mass 
excess (say) of $\delta M_{\lambda}^{{\rm rms}}$
over length scale $\lambda$, the most
obvious contribution to the CMB anisotropy in this
scale will be a perturbation in the gravitational 
potential of the form
\begin{equation} 
\delta\Phi_{\lambda} ({\bf r}) \approx 
\frac{G\delta M_{\lambda}}{\lambda} \approx
\frac{G \delta M_{\lambda}}{\lambda M_{\lambda}} \rho_0 V_{\lambda}.
\end{equation}
Since $V_{\lambda} \sim \lambda^3$ and from Eq. (8) we have
$\delta M_{\lambda}/M_{\lambda} \sim k_{\lambda}^2 \sim
\lambda^{-2}$ for $P(k) \sim k$, it is then apparent  that
\begin{equation}
\left(\frac{\delta T}{T}\right)_{\lambda} \approx  \delta\Phi_{\lambda}
= {\rm constant},
\end{equation}
i.e. for primordial HZ fluctuations the value of smoothed
temperature anisotropy is independent of the length scale
of the smoothing filter.  This is the well known Sachs-Wolfe
effect as applied to a HZ matter spectrum.  

Let us however query whether the standard result described above,
which concerns anisotropies caused by fluctuation in the mass surrounding
the site of photon emission and extending to the length scale 
$\lambda$ under consideration,
represent all the possibilities,  Of course, on scales
$\gg \lambda$ any total mass variation will, in the restricted context of
Eqs. (10) and (11), lead to anisotropies over correspondingly larger angular
separations.  Yet Eq. (10) is not the only way to perturb 
$\Phi_{\lambda} ({\bf r})$.  A simple analogy is that the potential at
some point on the earth surface depends solely on the total mass of the
earth if there is perfect spherical symmetry in the density function.
Deviations in $\Phi$ from point to point on the surface can occur even
when the total mass is fixed, if non-uniformities in the matter distribution
exists in {\it any} part of the earth's interior, far or near the
location in question.  Moreover, the distance over
which potential excursions occur equals the mean spacing
between mass concentrations.  In the present problem, obviously
beyond the distance $\lambda$
from a CMB emitter at position ${\bf r}$,
space remains just as inhomogeneous on the scale of $\lambda$.  Then,
different configurational realizations
(or random placement) of these size $\lambda$ density contrasts
(here-and-after referred to as lumps) within the,
horizon at decoupling, i.e. the causal sphere ${\cal R}$
of radius $R \gg \lambda$ centered at the emission point ${\bf r}$, can
also result in a finite $\delta\Phi_{\lambda} ({\bf r})$ which is unrelated
to Eq. (10).

Thus our contention is that when the distribution of
primordial matter is smoothed at resolution $\lambda$, the resulting
lumps of over- and underdensity, being
superposition states of many
independent Fourier modes with no preferred direction,
can randomly trade spaces with each other
and the
HZ nature of the matter spectrum
(on scales upwards of $\lambda$, of course) will still be preserved
\footnote{If the matter is
of primordial origin
the fluctuation in the total mass of
any multiple lump region
will remain consistent with the HZ spectrum, irrespective of how the
lumps are re-arranged,
provided that the volume filling factor (by the lumps) is
100 \%, and the procedure of re-configuration
ensures randomness of the lump positions within
some very large volume.}.  This
leads to an additional component of $\delta\Phi_{\lambda}({\bf r})$
from the configurational arbitrariness of an entire ensemble of lumps, all in
causal contact with position ${\bf r}$.

It is in fact quite easy
to estimate the magnitude of the additional contribution.  If primordial
lumps of various masses and size $\lambda$ randomly pack the Universe
with 100 \% filling factor, the resulting total mass fluctuation in any
multiple-lump sub-region will remain in compliance with
$\delta M/M \sim k^2$, yet there will be a perturbation on the potential
$\Phi_{\lambda} ({\bf r})$, given by
$\delta\Phi_{\lambda}^{{\rm rms}}/\Phi_{\lambda} \sim 1/\sqrt{N}$ 
where $N \approx
R^3/\lambda^3$ is the total number of lumps in our causal sphere, although
we shall derive the precise value of $\delta\Phi_{\lambda}^{{\rm rms}}$ in the
next section.  Thus we deduce that there is more CMB temperature
anisotropy to be reckoned with, of order
\begin{equation}
\left(\frac{\delta T}{T}\right)_{\lambda} \approx 
\delta\Phi_{\lambda}^{{\rm rms}}
\approx \frac{G\sqrt{N} \delta M_{\lambda}}{R},
\end{equation}
which has {\it not} been taken into account by conventional 
(single lump) application
of the Sachs-Wolfe effect, Eq. (11).
For a HZ matter spectrum, use of Eq. (8) 
enables us to write Eq. (12) as
\begin{equation} 
\left(\frac{\delta T}{T}\right)_{\lambda} \approx {\rm constant}~
\sqrt{\frac{R}{\lambda}},
\end{equation}
where the constant factor is the same as that in Eq. (11).
Bearing in mind that $R > \lambda$, the new effect presented here appears
at least as important as the standard result.

\vspace{2mm}

\noindent
{\bf 3. Anisotropy from analysis in configurational space}

A slightly different way of looking at the physics described in section 2 is
afforded by working in real space.  After spatial filtering the 
excursions in the potential, Eq. (2), can be written as
a discrete sum over
the resolution elements, viz.
\begin{equation}
\delta\Phi_{\lambda}({\bf r}) = \sum_i 
-G\frac{\delta M_{\lambda}^i}{|{\bf r} - {\bf r_i}|}.
\end{equation}
Here the mass fluctuation at position $i$ is given by
$\delta M_{\lambda}^i$, a quantity with zero mean and
standard deviation $\delta M_{\lambda}^{{\rm rms}}$ where
\begin{equation}
\delta M_{\lambda}^{{\rm rms}} = 
\frac{4}{3} \pi \lambda^3 \delta\rho_{\lambda}^{{\rm rms}}.
\end{equation}
Now the contribution to $\delta\Phi_{\lambda} ({\bf r})$
from the density contrast in the vicinity of the point
${\bf r}$, i.e. the `local lump' at ${\bf r_i} = {\bf r_0}$
where $|{\bf r} - {\bf r_0}| \sim \lambda$, is
\begin{equation}     
\delta\Phi_{\lambda}({\bf r}) = 
\frac{G \delta M_{\lambda}^{{\rm rms}}}{\lambda} = 
\frac{G}{\lambda}
\frac{\delta M_\lambda^{{\rm rms}}}{M_\lambda} \frac{4}{3} \pi\rho\lambda^3 = 
\frac{1}{2} \Omega_m H_0^2
\lambda^2 \frac{\delta\rho_{\lambda}^{{\rm rms}}}{\rho_\lambda},
\end{equation}
where $\Omega_m$ is the mean matter density of the present Universe
in units of the critical density $\rho_c = 3H_0^2/(8\pi G)$.
This yields
a temperature change (as the emitted CMB radiation leaves its own
region of potential excursion) of $\delta T/T \approx \delta\Phi_\lambda$, in
agreement with Eq. (48) of Sachs \& Wolfe (1967).  Note that because
in the case of a HZ spectrum
the quantity $\delta\rho_{\lambda}^{{\rm rms}}/\rho_\lambda 
\sim \delta M_{\lambda}^{{\rm rms}}/
M_{\lambda} \sim 1/\lambda^2$, we have from Eq. (8)
$\delta\Phi_{\lambda} ({\bf r}) =$ constant, in agreement with the
conclusion of the previous section that single lump Sachs-Wolfe effect leads
to the standard result of
scale invariant CMB anisotropy for primoridal matter distributions.

The important point, however, is that Eq. (14) also depicts effects
beyond that of the local lump.  As stated earlier, the contribution
to $\delta\Phi_{\lambda} ({\bf r})$ from the random placement all other 
lumps within
the horizon should be included with
the summation procedure.  It is emphasized again that this effect is
{\it not because of the Poisson statistics in the total mass of many
independently varying density contrasts}
belonging to the causal sphere ${\cal R}$
(which is suppressed by the HZ spectrum of $P(k) \sim k$), but
because of the Poisson process in the configurational arrangement
of the {\it same} primordial lumps inside ${\cal R}$.

Our task initially is to compute the deviation
$\delta\Phi_{\lambda}^{{\rm rms}}$ in the
gravitational potential of a sphere of radius $R$ when it contains a 
{\it constant} number 
$N$ of smaller spherical lump of radius $\lambda$, each randomly
placed and filled with matter
accounting for a mass of $\delta M_{\lambda}^{{\rm rms}}$ per lump,
such that the remaining matter-free
regions occupy half the volume of the causal sphere.  Next, it is realized that
this actually means neglecting the contribution to 
$\delta\Phi_{\lambda}^{{\rm rms}}$ from
the underdense regions (each of mass $-\delta M_{\lambda}^{{\rm rms}}$),
when it is included $\delta\Phi_{\lambda}^{{\rm rms}}$ will
increase by a factor of two because the gravitational effects of
these two types of regions are anti-correlated.
The potential deviation at the position $\bf r$ ($r \leq R$) due
to the overdense lumps is
 \begin{equation}
\delta\Phi_{\lambda}({\bf r}) =
-\sum_j\frac{G\delta M_{\lambda}^{{\rm rms}}}{|{\bf r} - {\bf r_j}|}.
 \end{equation}
Now each of the $N$ overdense lumps
has the same probability distribution, in which the probability
$p({\bf r})d^3 {\bf r}$ of finding the lump in a small
volume $d^3 {\bf r}$ is given by
 \begin{equation}
p({\bf r})d^3 {\bf r}=\frac{n({\bf r})}{N}d^3 {\bf r}.
 \end{equation}
where in Eq. (18) we adopted the continuum approximation
appropriate to the limit of many lumps.
Hence the mean potential $\langle\Phi ({\bf r}) \rangle$ is
 \begin{equation}
\langle\delta\Phi_{\lambda}({\bf r})\rangle=-NG\delta M_{\lambda}^{{\rm rms}} 
\left\langle\frac{1}{|{\bf r} - {\bf r'}|}
\right\rangle=-G\delta M_{\lambda}^{{\rm rms}}
\int\frac{n({\bf r'})d^3 {\bf r'}}{|{\bf r} - {\bf r'}|}.
 \end{equation}  
Relevant to the present problem is a fixed set
of overdense lumps within
${\cal R}$, i.e. $n({\bf r}) = 3N/(4\pi R^3)$ 
for $r \leq R$, and $n({\bf r}) = 0$
for $r > R$ since the matter lying beyond ${\cal R}$ plays no part.
The integral of
Eq. (19) can readily be evaluated under this scenario to yield
 \begin{equation}
\langle\delta\Phi_{\lambda}({\bf r})\rangle = 
-\frac{GN\delta M_{\lambda}^{{\rm rms}}}{2R} \left( 3 - \frac{r^2}{R^2}
\right), 
 \end{equation}  
where Eq. (20) is valid for the range $0 < r \leq R$.  Thus at $r=R$ one
recovers the usual Newtonian potential $\langle\delta\Phi_{\lambda}\rangle = 
-GN\delta M_{\lambda}^{{\rm rms}}/R$, which on
average cancels exactly the contribution from the underdense lumps.

To obtain the variance 
$\langle(\delta \Phi_{\lambda})^2\rangle$, the quantity by which
the potential for points in space separated by
lengths $> 2\lambda$ differ from each other,
we need the mean square of $\delta\Phi_{\lambda}$
defined as
 \begin{equation}
\langle[\delta\Phi_{\lambda} ({\bf r})]^2 \rangle=\sum_{j,k}
\left\langle
\frac{G^2 (\delta M_{\lambda}^{{\rm rms}})^2}{|{\bf r} - {\bf r_j}||{\bf r} 
- {\bf r_k}|}\right\rangle.
 \end{equation}
When we subtract from this the `square of the mean', viz. the quantity
 \begin{equation}
\langle\delta\Phi_{\lambda} ({\bf r})\rangle^2 = 
\left\langle\sum_j\frac{G\delta M_{\lambda}^{{\rm rms}}}{|{\bf r} - {\bf r_j}|}
\right\rangle^2,
 \end{equation}
it is clear that all the terms with
$j\ne k$ will cancel, and we are left with
 \begin{equation}
\langle[\delta\Phi_{\lambda} ({\bf r})]^2\rangle-
\langle\delta\Phi_{\lambda} ({\bf r})\rangle^2=
G^2 (\delta M_{\lambda}^{{\rm rms}})^2 N
\left(\left\langle\frac{1}{|{\bf r} - {\bf r'}|^2}\right\rangle-
\left\langle\frac{1}{|{\bf r} - {\bf r'}|}\right\rangle^2\right).
 \end{equation}
Now the average related to the first term on
the right side of Eq. (23) may also be computed analytically for the
form of $n({\bf r})$ already discussed.  By this we mean
 \begin{equation}
\left\langle\frac{1}{|{\bf r} - {\bf r'}|^2}\right\rangle  = \frac{3}{2R^3}
\int_0^r \frac{r'dr'}{r} \ln\left(\frac{r+r'}{|r-r'|}\right)
 \end{equation}
At the surface itself $r=R$, and the integral reduces to the simple form
 \begin{equation}
\left\langle\frac{1}{|{\bf r} -
{\bf r'}|^2}\right\rangle = \frac{3}{2R^2}~
{\rm at}~r=R.
 \end{equation}
The 2nd term on the right side of Eq. (23) is, by Eq. (19),
equal to $\langle\delta\Phi_{\lambda}\rangle^2/N$.
 
Thus, altogether and after including the aforementioned
contribution to $\delta\Phi_{\lambda}^{{\rm rms}}$
from the underdense lumps, we arrive at a standard deviation of
 \begin{equation}
\delta\Phi_{\lambda}^{{\rm rms}} ({\bf r}) = 
2\left[ \frac{3G^2 (\delta M_{\lambda}^{{\rm rms}})^2 N}{2R^2} -
\frac{G^2 (\delta M_{\lambda}^{{\rm rms}})^2 N}{R^2} \right]^{\frac{1}{2}} =
\frac{\sqrt{2N} G \delta M_{\lambda}^{{\rm rms}}}{R},
 \end{equation}
Eq. (26) depicts the spatial variation in the potential between
two points located at a distance $\geq 2\lambda$ apart.

The CMB anisotropy is finally obtained from Eq. (26) by observing that
within the causal sphere there are altogether $R^3/\lambda^3$ lumps
of radius $\lambda$, half of which are overdense.  Thus
$N = R^3/(2\lambda^3)$, and we have
\begin{equation}
\left(\frac{\delta T}{T}\right)_{\lambda} 
\approx \delta\Phi_{\lambda}^{{\rm rms}} =
G \delta M_{\lambda}^{{\rm rms}} \lambda^{-\frac{3}{2}} R^{\frac{1}{2}} =
\frac{1}{2} \Omega_m H_0^2
\lambda^{\frac{3}{2}} R^{\frac{1}{2}}
\frac{\delta\rho_{\lambda}^{{\rm rms}}}{\rho_{\lambda}}~{\rm for}~
\lambda < R.
 \end{equation}
This gives the new contribution
which should {\it explicitly} be invoked\footnote{That is to say, the
term will not arise `naturally' as solution of a large number of
coupled differential equations unless the {\it physics} behind it is
in place.} as
a separate anisotropy term in the `Boltzmann solver' codes like
CMBFAST (the routine used to fit
the standard cosmological model to WMAP data, see
Zaldarriaga \& Seljak 2000), but
is not.  More elaborately,
for self-consistency of the standard model this
contribution must
be included, even though it has hitherto been ignored.
When compared with Eq. (16), we see that although conventional Sachs-Wolfe
effect yields a constant $\delta T/T$ at all scales, the extra
anisotropy $(\delta T/T)_{\lambda}$ 
equals this same constant multiplied by the factor $\sqrt{R/\lambda}$
(in agreement with the earlier analysis of section 3),
i.e. it is {\it not} scale invariant.  The reason why CMB temperatures
must exhibit the variation of Eq. (27) is that two points on the last
scattering surface separated by a distance $\geq 2\lambda$ experience
statistically independent potential deviation, due to the
different realization (or pattern) of lumps in the causal spheres
centered at these points.  

It should also be mentioned
that no matter how many more
contrainsts unassumed by us were to be applied to restrict the number
of permitted lump configurations within ${\cal R}$,
so long as space is not totally homogeneous
beyond the emitter's own `local lump' the incorporation of effects of
remote density contrasts via the long range force of gravity {\it must}
remove the constancy of $(\delta T/T)_{\lambda}$ by introducing a
factor that scales with $R/\lambda$ in some way.  The fact that the
standard model predicts a constant $(\delta T/T)_{\lambda}$ at large
(but sub-horizon) $\lambda$ offers the clearest indication of it's
failure in recognizing the phenomenon presented here.

\vspace{2mm}

\noindent
{\bf 4.  Have the CMB anisotropy observations been interpreted correctly?}

At sufficiently low spherical harmonics $\ell <$ 90 where
the smoothing length $\lambda > R$, lack of
causal linkage between lumps renders Eq. (27) inapplicable, and
the only anisotropy would be the conventional single
lump Sach-Wolfe effect of Eq. (16).  From the WMAP TT cross correlation
plot (Bennett et al 2003) one sees that $(\delta T/T)_{\lambda} 
\approx 10^{-5}$ at $\ell \ll$ 90.  Thus we conclude that
\begin{equation}
\frac{1}{2} \Omega_m H_0^2
\lambda^2
\frac{\delta\rho_{\lambda}^{{\rm rms}}}{\rho_{\lambda}} \approx 10^{-5}
\end{equation}
on all scales $\lambda$ so long as the matter distribution follows
the primordial HZ spectrum of $P(k) \sim k$.  

For sub-horizon features 
at decoupling which were super-horizon in size during matter-radiation
equipartition\footnote{Structures that fit inside the comoving 
equipartition horizon
have according to the standard model a density contrast commensurate with 
$P(k) \sim k^{-3}$ at the primordial level.  
Thus our present conclusion regarding the first
peak, which is however based upon a $P(k) \sim k$ spectrum, 
does {\it not} apply to the higher peaks.},
like the first acoustic peak,
the extra anisotropy of Eq. (27)
must now be included.  Since $R \approx$ 490 Mpc
(present value for the causal radius at decoupling) and, in the
case of the first peak the full size of these lumps is
$2\lambda \approx$ 147 Mpc, we see from Eq. (27) and (28) that the
total expected anisotropy with the conventional and new component
added in quadrature is
\begin{equation}
\frac{\delta T}{T} \approx 2.77 \times 10^{-5}
\end{equation}
for the first peak.   Obviously, Eq. (29) in conjunction
with the WMAP observations pose a dilemma for
the standard  
cosmological model,
because the detected first
acoustic anisotropy is only $\approx$ 2.5 $\times$ 10$^{-5}$, it
already equals
the expected value as given by Eq. (29), which is based upon
a purely primordial spectrum {\it before enhancement of density
contrasts by sound waves}.  This leaves no room for interpreting
the acoustic features as compression and rarefaction of the
plasma by sound propagation.

Is  there a way of restoring the standard cosmological model,
which matches the WMAP data very well?  We could
supposedly alter the cosmological
parameters, e.g. the primordial spectral index could substantially
increase from $n=1$ to reduce the impact of Eq. (27), though the
epistimological justification of such changes will almost
certainly be artificial.  One less damaging possibility remains, however.
The gravitational force beyond some $>$ 100 Mpc distance scale could
wane more quickly than the inverse-square law.  This would appear
to the author as a remedy of `least resistance' for several reasons.
Firstly, the standard model can immediately be reinstated, as the
Friedmann equations are constructed
in the context of General Relativity without appealing to
the existence
of a gravitational influence on Hubble scales.  Secondly, structure formation
theory is unaffected, because numerical codes that simulate the building
galaxies, groups, and even clusters do not depend on the inverse-square
law persisting to beyond 100 Mpc.  Thirdly, there is currently no reliable
experimental constraint on the behavior of gravity at these very
large distances. 

\vspace{2mm}

\noindent
{\bf References}

\noindent
Bennett, C.L. et al, 2003, ApJS, 148, 1-27. 

\noindent
Bond, J.R., \& Efstathiou, 1984, ApJ, 285, L45.

\noindent 
Bond, J.R., \& Efstathiou, 1987, MNRAS, 226, 655.

\noindent
Harrison, E.R., 1970, MNRAS, 147, 279.

\noindent
Page, L. et al 2003, ApJS, 148, 233.

\noindent
Peebles, P.J.E., \& Yu, J.T., 1970, ApJ, 162, 815.

\noindent
Peebles, P.J.E., 1982, ApJ, 263, L1.

\noindent
Sachs, R.K., \& Wolfe, A.M., 1967, ApJ, 147, 73.

\noindent
Spergel, D.N. et al, 2003, ApJS, 148, 175.

\noindent
Zaldarriaga, M., \& Seljak, U. 2000, ApJS, 129, 431.

\noindent
Zel'dovich, Ya. B., 1972,  MNRAS, 160, 1.

\end{document}